\documentclass{emulateapj}
\usepackage{graphicx}
\usepackage{natbib}
\usepackage{multirow}
\def\rmsq {rad\,m$^{-2}$}

\def\etal {\textit{et al.}}

\slugcomment{accepted for publication in ApJ}

\shorttitle{Faraday Rotation Toward 3C\,84}
\shortauthors{Plambeck \etal}

\begin{document}

\title{Probing the Parsec-Scale Accretion Flow of 3C\,84 with Millimeter Polarimetry}


\author{R.~L.~Plambeck\altaffilmark{1},
G.~C.~Bower\altaffilmark{2},
Ramprasad~Rao\altaffilmark{2},
D.~P.~Marrone\altaffilmark{3},
S.~G.~Jorstad\altaffilmark{4,5},
A.~P.~Marscher\altaffilmark{4},
S.~S.~Doeleman\altaffilmark{6,7},
V.~L.~Fish\altaffilmark{6}, 
M.~D.~Johnson\altaffilmark{7}}

\altaffiltext{1}{Radio Astronomy Laboratory, University of California, Berkeley, CA 94720, USA}
\email{plambeck@berkeley.edu}
\altaffiltext{2}{Academia Sinica Institute for Astronomy and Astrophysics (ASIAA), 645 N. Aohoku Pl., Hilo, HI 96720, USA}
\altaffiltext{3}{Steward Observatory, University of Arizona, 933 North Cherry Avenue, Tucson, AZ 85721, USA}
\altaffiltext{4}{Institute for Astrophysical Research, Boston University, 725 Commonwealth Avenue, Boston, MA 02215, USA}
\altaffiltext{5}{Astronomical Institute, St. Petersburg State University, Universitetskij Pr. 28, 198504 St. Petersburg, Russia}
\altaffiltext{6}{MIT Haystack Observatory, Route 40, Westford, MA 01886, USA}
\altaffiltext{7}{Harvard-Smithsonian Center for Astrophysics, 60 Garden Street, Cambridge, MA 02138, USA}

\begin{abstract}

We report the discovery of Faraday rotation toward radio source
3C\,84, the active galactic nucleus in NGC1275 at the core of the Perseus
Cluster.  The rotation measure (RM), determined from polarization observations
at wavelengths of 1.3 and 0.9\,mm, is ($8.7 \pm 2.3$)$\times 10^5$ \rmsq, among
the largest ever measured.  The RM remained relatively constant over a 2 year
period even as the intrinsic polarization position angle wrapped through a span
of 300 degrees.  The Faraday rotation is likely to originate either in the
boundary layer of the radio jet from the nucleus, or in the accretion flow onto
the central black hole.  The accretion flow probably is disk-like rather than
spherical on scales of less than a parsec, otherwise the RM would be even
larger.

\end{abstract}

\keywords{accretion, accretion disks --- polarization --- galaxies: active ---
galaxies:  jets --- galaxies:  individual(3C\,84)}

\section{INTRODUCTION}
\label{sec:intro} 

Radio source 3C\,84 is associated with the active galactic nucleus (AGN) in
NGC\,1275, the central galaxy in the Perseus cluster, the prototypical `cooling
flow' cluster \citep{Fabian1994}.  The black hole in the AGN launches powerful
jets into the surrounding medium.  The accretion process onto the black hole
has been studied through a variety of techniques on scales as small as a few
parsecs \citep{Vermeulen1994, Walker1994, Walker2000, Wilman2005,
Scharwachter2013}.  At the distance of NGC1275, 1 pc subtends 3
milliarcseconds.

At cm wavelengths 3C\,84 is well-known as an `unpolarized' calibrator.  Why is
this so, given that the radio emission from the AGN and its associated jet
arise from synchrotron emission, which should be highly polarized?  One
possibility is that Faraday rotation twists the position angle $\chi$ of this
linearly polarized radiation as it propagates through foreground plasma.  The
position angle is rotated by $\Delta\chi = {\rm RM}\,\lambda^2$, where RM is
the rotation measure.  If RM varies across the source and the observations do
not resolve this structure (``beam depolarization''), the net observed
polarization may be very small.  

Measurements of Faraday rotation along the line of sight to the black hole
provide a valuable diagnostic of the accretion flow onto the central object,
since the RM is proportional to the integral of the electron density and the
magnetic field along the line of sight.  In the case of SgrA*, for example, the
RM has been used to constrain both the mode and the rate of the accretion onto
its black hole \citep{Bower2003,Marrone2007}.  Similar methods have recently
been applied to M87 \citep{Kuo2014}.  Time variability of the RM could also be
a valuable probe of turbulence in the accretion region \citep{Pang2011}.

For 3C\,84, \citet{Taylor2006} found an RM of about 7000~\rmsq\ toward a small
spot in the jet about 15 milliarcseconds ($\sim5$~pc) south of the nucleus,
based on VLBA maps at wavelengths of 1.3, 2.0, and 3.6~cm. It was not possible
to fit the RM toward the nucleus itself because in that direction linear
polarization was detected only at a single wavelength (and only at the 0.2\%
level).  At 7\,mm, where emission from the nucleus becomes dominant, VLBA
monitoring observations by the Boston University
group\footnote{http://www.bu.edu/blazars/VLBAproject.html} \citep{Marscher2011}
sometimes detect spots of weak linear polarization toward the nucleus, but
typically the polarized flux density is $< 0.5$\% of the peak flux density.

Polarization should be easier to detect at mm wavelengths because Faraday
rotation decreases steeply at shorter wavelengths, and because the mm emission
region is smaller, so that variations in RM across the source are less
problematic.  However, based on observations made with the Plateau de Bure
interferometer in 2011 Mar, \citet{Trippe2012} placed upper limits of 0.5\% on
the linear polarization of 3C\,84 at wavelengths of 1.3 and 0.9 mm.  Here we
report observations at the same wavelengths made over a 2 year period with the
Combined Array for Research in Millimeter Astronomy (CARMA) and with the
Submillimeter Array (SMA).  The fractional polarization of 3C\,84 was $< 0.6\%$
in the earliest data, from 2011 May, consistent with the \citet{Trippe2012}
results, but by late 2011 it had increased to the 1--2\% level.  The RM
inferred from the data is $\sim 9 \times 10^5$~\rmsq, among the largest ever
measured.  We discuss the implications of these results for the accretion flow
onto the black hole in 3C\,84.

\vspace{12pt}

\section{OBSERVATIONS}
\label{sec:obs}

\subsection{CARMA Observations}

The CARMA polarization system \citep{Hull2013,Hull2014} consists of
dual-polarization 1.3\,mm receivers that are sensitive to right- (R) and
left-circular (L) polarization, and a spectral-line correlator that measures
all four cross-correlations  (RR, LL, LR, RL) on each of the 105 baselines
connecting the 15 antennas.  

The double sideband receivers are sensitive to signals  at sky frequencies
$\nu_{\rm sky} = \nu_{\rm LO} \pm \nu_{\rm IF}$ above (upper sideband) and
below (lower sideband) the local oscillator frequency $\nu_{\rm LO}$. Signals
received in these two sidebands are separated in cross-correlation spectra.
The correlator provides 4 independently tunable sections, each up to 500~MHz
wide.  Typically we centered these sections at intermediate frequencies
$\nu_{\rm IF}$ of 6--8 GHz, so that the polarization data from the upper and
lower sidebands spanned a sky frequency range of 16~GHz.

Data were analyzed with the MIRIAD package \citep{Sault1995}.  Stokes
parameters $I$, $Q$, and $U$ were derived for each of the 8 spectral windows (4
sections $\times$ 2 sidebands).  $Q$ and $U$ may be considered components of a
complex polarization vector $p = Q+iU = p_0\, {\rm exp}(i2\chi)$. Here $p_0$ is
the linearly polarized flux density in Jy, $\chi(\nu) = \chi_0 + {\rm RM}
(c^2/\nu^2-c^2/\nu_0^2)/(1+z)^{2}$ is the electric vector position
angle, RM is the rotation measure, $z$ is the redshift, and $\chi_0$ is the
position angle at the reference frequency $\nu_0$ (225~GHz). The factor
$(1+z)^2$ arises because Faraday rotation takes place at frequency
$\nu(1+z)$ in the source frame; this correction factor is negligible for
3C\,84, at $z=0.018$.  We fit $Q(\nu)$ and $U(\nu)$ to solve for $p_0$,
$\chi_0$, and RM.

For a bright but weakly polarized source like 3C\,84, the accuracy of the
measurements is limited by systematic errors, not thermal noise.  The primary
difficulty is in correcting for the polarization leakages -- the cross coupling
between the L and R channels caused by imperfections in the receivers or
crosstalk in the IF system.  Leakages are derived from observations of a bright
point source, polarized or unpolarized (usually 3C\,84 itself), obtained over a
wide range of parallactic angle.  Miriad task {\tt gpcal} fits these data to
solve simultaneously for the source polarization, receiver gains, and leakage
corrections.  Since the CARMA receivers have no moving parts the leakages are
stable over periods of months.  Their magnitudes are typically of order 6\%,
however, and they have considerable frequency structure.  We calibrated the
leakages separately for each of the 8 spectral windows.  

We were able to set only a crude upper limit of \mbox{$\lesssim 2$\%} on the
magnitude of circular polarization (Stokes $V$) because this requires highly
accurate calibration of the gains of the R and L channels on a source other
than 3C\,84.

\subsection{SMA Observations}

SMA observations were conducted in both the 1.3\,mm and 0.9\,mm bands.  The
single polarization receivers are switched between R and L circular
polarization by inserting quarter wave plates into the optical path.  Using a
different switching pattern for each of the 8 telescopes, all 4
cross-polarizations (RR, LL, LR, RL) are measured every 5 minutes on each
baseline.  Like CARMA, the SMA operates in double sideband mode, with a 4--8
GHz IF.  The available observational bandwidths were either 2 GHz or 4 GHz.
Thus the data spanned a sky frequency range of either 10~GHz or 12~GHz.

Data were reduced using a combination of the MIR/IDL and MIRIAD data reduction
packages.  The instrumental polarization is frequency-dependent and the typical
values are $\sim 2\%$.  The instrumental polarization is determined with
an accuracy of $\sim0.1\%$.  The RM was fit to the difference in the upper and
lower sideband position angles.

\section{RESULTS}
\label{sec:results}

\begin{figure} \centering
\includegraphics[width=1.0\columnwidth, clip, trim=1.5cm 5cm 1cm 5cm] {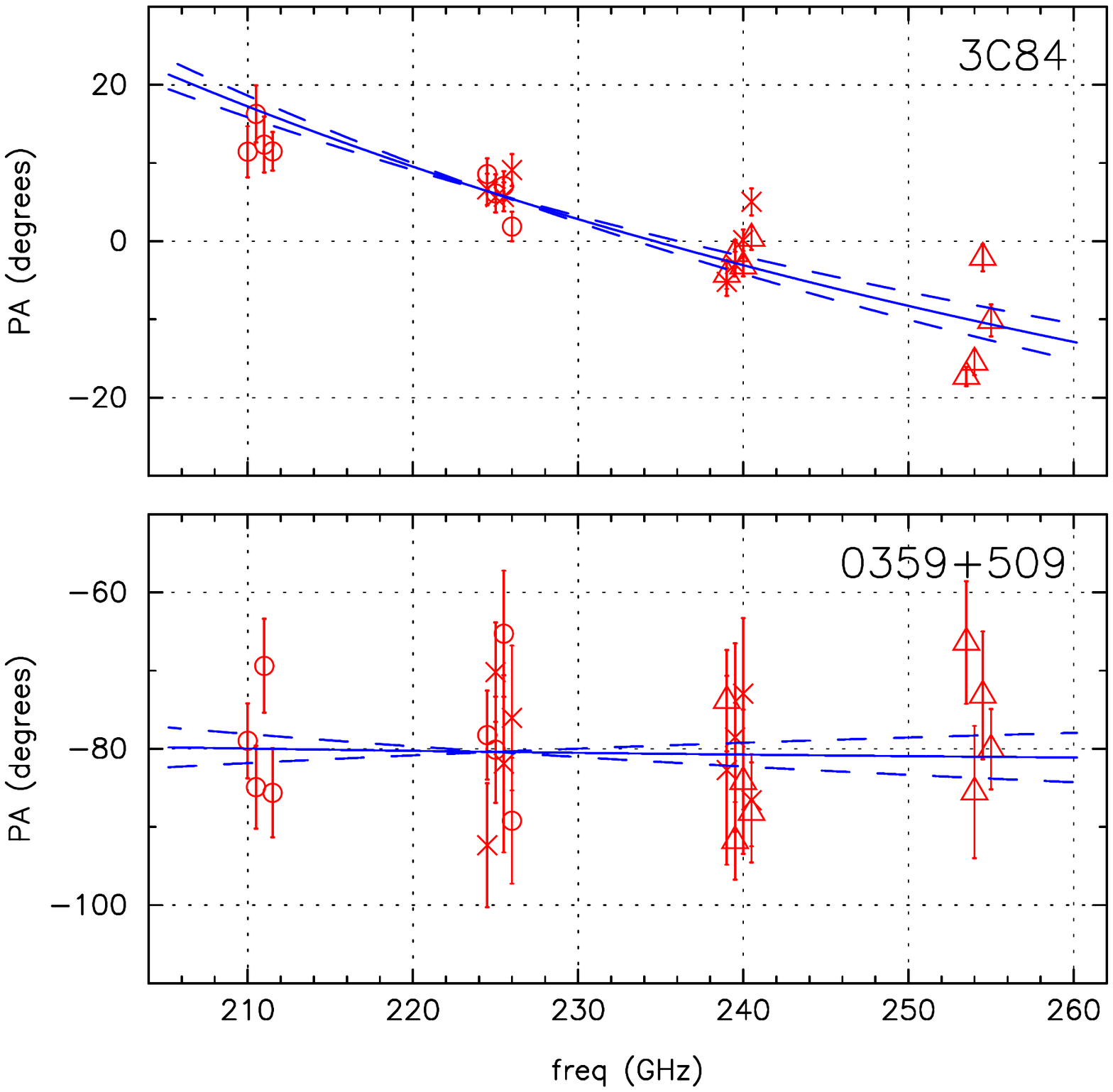} 

\caption{Polarization position angles from 211 to 255 GHz for 3C\,84 and
comparison source 0359+509 measured with CARMA on 2013 Aug 04.  Symbols
indicate the LO frequency used for the observations -- circles, 218~GHz;
crosses, 232.5~GHz; triangles, 247~GHz.  The sky frequencies observed at
$\nu_{\rm LO}$ of 232.5~GHz overlap those observed at 218 and 247~GHz. The
error bars are estimated from the scatter in the measurements at each frequency
and do not fully reflect systematic errors in the polarization leakage
calibration.  Fractional polarizations of 1.5\% were measured for both 3C\,84
and 0359+509; 0359+509 has a much lower flux density than 3C\,84, so the
position angle uncertainties due to thermal noise are larger.  Rotation
measures are derived from fits to the position angle vs. frequency, indicated
by the blue curves.} 

\label{fig:PA04aug2013} 
\vspace{12pt} 
\end{figure}

The 3C\,84 data reported here span the period from 2011 May through 2013
August.  In almost all cases 3C\,84 was observed as a calibrator for another
science target.  Many of the CARMA datasets were from the TADPOL survey
\citep{Hull2014}.

One CARMA observation targeted 3C\,84 specifically.  In an 8-hour observation
on 2013 August 04 we interleaved observations at LO frequencies of 218, 232.5,
and 247~GHz to obtain wide parallactic angle coverage at  16 sky
frequencies from 210--255~GHz. Both 3C\,84 and a comparison calibrator,
0359+509, were observed. Fits to these data, shown in
Figure~\ref{fig:PA04aug2013}, give RM of  \mbox{($7 \pm 1$)$\times 10^5$~\rmsq}
for 3C\,84  and \mbox{$(1.9\pm 7.6)\times 10^5$~\rmsq} for
0359+509.  The uncertainty is large for 0359+509 because this source is at
redshift $z$=1.52, and the RM scales as $(1+z)^2$; however, the
0359+509 data rule out the possibility that the systematic 25$\arcdeg$ position
angle variation measured for 3C\,84 could be an instrumental effect.

\begin{deluxetable}{lccrr} 
\tablecaption{CARMA and SMA Observations of 3C\,84}
\tablehead{
\colhead{Epoch} & 
\colhead{$\nu_{\rm LO}$} & 
\colhead{$p$\tablenotemark{a}} & 
\colhead{$\chi$\tablenotemark{b}} & 
\colhead{RM} \\ 
                 &  (GHz) & (\%) & (deg) & ($10^5$ rad m$^{-2}$) 
} 
\startdata 
\multicolumn{5}{c}{CARMA 1.3\,mm} \\ 
\hline 
   2011May03 \rule{0in}{0.11in} & 223.8 & 0.6 & $  -63 \pm    8 $ & $ -14.0 \pm  19.0 $ \\ 
   2011Oct27 & 223.8 & 1.2 & $  -56 \pm    2 $ & $   9.8 \pm   3.0 $ \\ 
   2011Nov09 & 223.8 & 1.3 & $  -59 \pm    2 $ & $   7.9 \pm   2.9 $ \\ 
   2012Apr07 & 223.8 & 1.2 & $  -76 \pm    3 $ & $   9.1 \pm   3.7 $ \\ 
   2012Jun24 & 223.8 & 1.5 & $   40 \pm    1 $ & $  11.0 \pm   2.1 $ \\ 
   2012Jul30 & 223.8 & 1.0 & $   13 \pm    2 $ & $  14.1 \pm   2.4 $ \\ 
   2012Sep02 & 223.8 & 1.5 & $  -13 \pm    1 $ & $  10.6 \pm   1.2 $ \\ 
   2012Oct18 & 223.8 & 1.0 & $  -41 \pm    2 $ & $   8.5 \pm   3.0 $ \\ 
   2012Oct30 & 223.8 & 0.8 & $  -49 \pm    4 $ & $   8.9 \pm   4.7 $ \\ 
   2012Nov24 & 223.8 & 1.2 & $  -86 \pm    2 $ & $   7.0 \pm   3.1 $ \\ 
   2013Mar22 & 226.3 & 1.4 & $   43 \pm    1 $ & $  10.6 \pm   2.3 $ \\ 
   2013Mar23 & 226.3 & 1.4 & $   29 \pm    2 $ & $   8.1 \pm   4.3 $ \\ 
   2013Aug04 & 232.5 & 1.5 & $    5 \pm    1 $ & $   7.0 \pm   0.9 $ \\ 
\hline 
\multicolumn{5}{c}{\rule{0in}{.11in} SMA 1.3\,mm} \\ 
\hline 
   2012Jun24 \rule{0in}{.11in} & 224.9 & 2.5 & $   34 \pm    1 $ & $   7.2 \pm   1.7 $ \\ 
   2012Jul20 & 226.9 & 1.4 & $   12 \pm    1 $ & $   7.6 \pm   2.7 $ \\ 
   2012Sep07 & 224.9 & 1.8 & $  -14 \pm    1 $ & $   5.9 \pm   2.1 $ \\ 
   2013Jan23 & 225.3 & 1.5 & $   73 \pm    1 $ & $   3.7 \pm   2.6 $ \\ 
   2013Jul05 & 226.9 & 3.2 & $   61 \pm    1 $ & $  10.5 \pm   1.2 $ \\ 
   2013Aug15 & 226.9 & 1.4 & $   12 \pm    1 $ & $   8.4 \pm   3.0 $ \\ 
\hline 
\multicolumn{5}{c}{ \rule{0in}{.11in} SMA 0.9\,mm} \\ 
\hline 
   2011Aug20 \rule{0in}{.11in} & 341.7 & 2.2 & $   85 \pm    3 $ & $   6.4 \pm  20.6 $ \\ 
   2012Jun15 & 343.0 & 2.0 & $    4 \pm    2 $ & $  16.3 \pm  13.8 $ \\ 
   2012Jul03 & 340.1 & 1.5 & $   -9 \pm    1 $ & $   0.0 \pm   8.9 $ \\ 
   2012Aug08 & 340.1 & 1.4 & $  -38 \pm    2 $ & $   9.4 \pm  15.7 $ \\ 
   2012Sep02 & 340.8 & 2.0 & $  -46 \pm    3 $ & $   3.2 \pm  17.9 $ \\ 
   2012Oct14 & 341.4 & 1.5 & $  -67 \pm    1 $ & $   9.6 \pm   9.1 $ \\ 
   2013Feb01 & 341.6 & 0.6 & $   32 \pm    4 $ & $ -22.5 \pm  22.8 $ \\ 
   2013Aug25 & 341.6 & 1.4 & $  -20 \pm    2 $ & $  -9.7 \pm  11.6 $ \\ 
\enddata 
\tablenotetext{a}{Fractional polarizations were {\it not} corrected for
noise bias, since the polarized flux density was typically more than
10 times the thermal noise level.}
\tablenotetext{b}{Polarization position angles $\chi$ are interpolated to
225 GHz for the 1.3\,mm data, and to 341~GHz for the 0.9\,mm data.}
\end{deluxetable} 

\begin{figure}
\centering
\includegraphics[width=1.0\columnwidth, clip, trim=1.5cm 2.55cm 1cm 3.2cm] {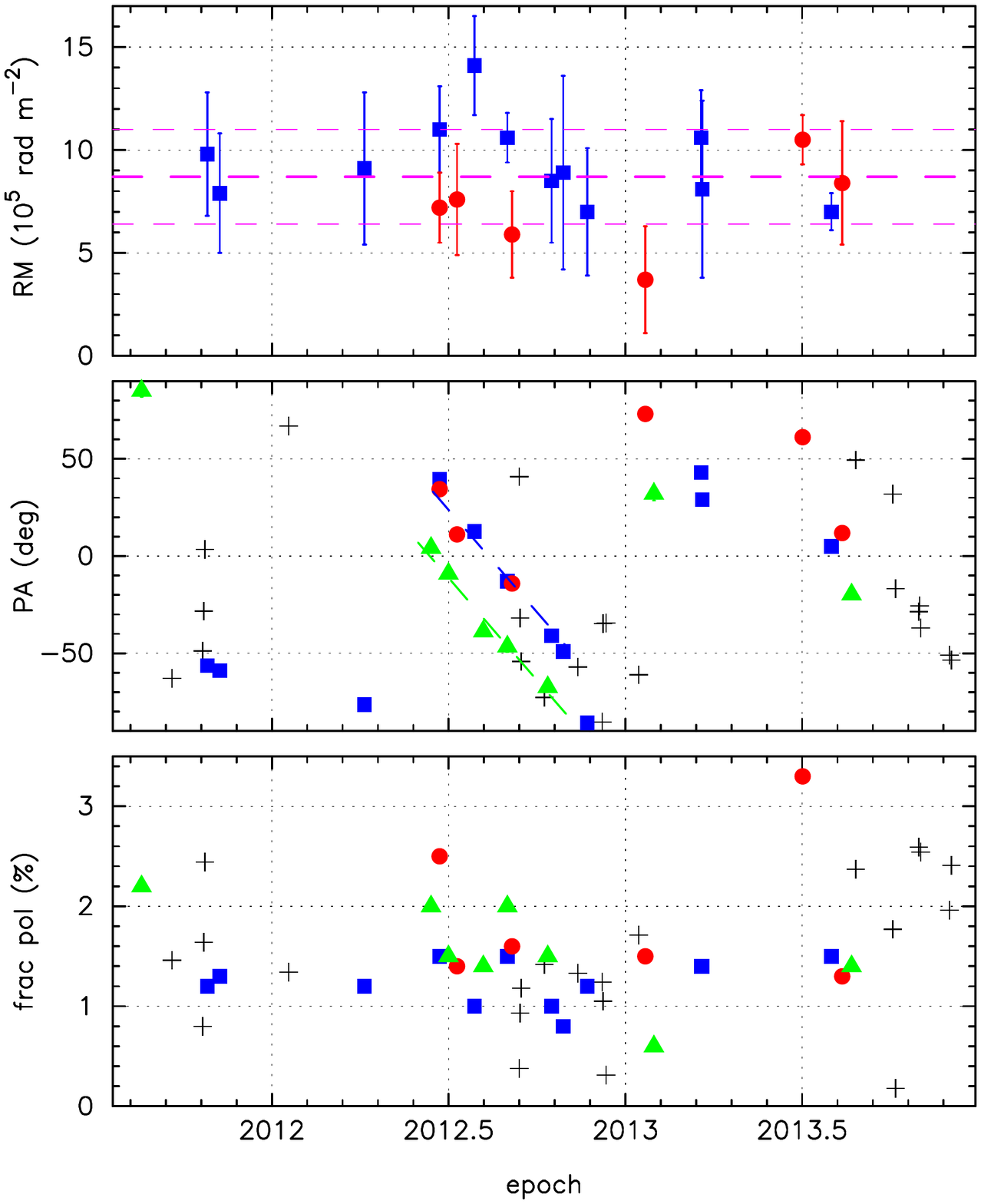}
\caption{Polarized intensities, electric vector position angles, and rotation
measures observed at 1.3\,mm (blue squares, CARMA; red circles, SMA) and
0.9\,mm (green triangles, SMA) from 2011 Aug through 2013 Dec.  Optical
fractional polarizations and position angles measured at Lowell Observatory are
shown by black crosses.  The dashed lines in the middle panel show that in
mid-2012 the position angles at 0.9\,mm were roughly $35\arcdeg$ more negative
than those at 1.3\,mm, consistent with a rotation measure of $6 \times
10^5$~\rmsq.  Dashed lines in the top panel show the mean $\pm$ 1 standard
deviation of the 1.3\,mm RM measurements.}
\label{fig:bigplot}
\vspace{0.05in}
\end{figure}

The fractional polarizations, position angles, and rotation measures derived
from all observations are summarized in Table 1 and plotted in
Figure~\ref{fig:bigplot}.  In our earliest data, from 2011 May, the fractional
polarization of 3C\,84 was very low, $\lesssim 0.6$\%, but for most of the
following observations it was in the 1--2\% range.  The polarization position
angle trended monotonically toward more negative values, apparently wrapping
through from $-90\arcdeg$ to $+90\arcdeg$ twice over the 2 year span of the
observations.

Also plotted in Figure~\ref{fig:bigplot} are the R-band optical polarizations
and position angles for 3C\,84 measured with the 1.8-m Perkins telescope at
Lowell Observatory (Flagstaff, AZ) using the PRISM camera.  The observations
and data reduction were performed in the same manner as described by
\citet{Jorstad2010} for the quasar 3C454.3.  The 1--2\% fractional polarization
in the optical is similar to that at mm wavelengths, but there is not a simple
correspondence between the optical and mm position angles; $\chi_{opt}$
sometimes fluctuates by tens of degrees on time scales of days, while
$\chi_{mm}$ tends to vary more smoothly.

Generally there is good agreement between the CARMA and SMA results at 1.3\,mm;
significant RM measurements were made with both instruments.  It was not
possible to detect Faraday rotation from the 0.9\,mm data alone -- at this
wavelength the expected position angle difference between the upper and lower
sidebands is only 2.5$\arcdeg$ for an RM of $10^6$~\rmsq.  However,
interpolation of the data in Figure~\ref{fig:bigplot} shows that position
angles at 0.9\,mm were 30--40$\arcdeg$ more negative than those at 1.3\,mm in
mid-2012.  The fact that this offset is maintained over a period of months even
as the position angles at both wavelengths rotate through 90 degrees provides
powerful evidence that we are observing Faraday rotation in an external screen,
rather than variations in the polarization direction vs.  synchrotron optical
depth.  A 35$\arcdeg \pm 5\arcdeg$ difference in the 1.3\,mm and 0.9\,mm
position angles corresponds to an RM of ($6 \pm 1$)$\times 10^5$~\rmsq.

The uncertainties in the RM measurements listed in Table~1 do not fully account
for possible systematic errors in the polarization leakage corrections.  Thus,
although our results allow for possibility of up to 50\% variations in the RM
on time scales of days or weeks, the evidence for such variations is not
convincing.  An average of the 1.3\,mm RM values, excluding the anomalous
result from 2011 May, gives \mbox{RM = ($8.7 \pm 2.3$)$\times 10^5$~\rmsq},
where the uncertainty is the standard deviation of the measurements.

\section{Interpretation}
\label{sec:discussion}

Where do the linearly polarized emission and Faraday rotation originate in
3C\,84, and what conclusions can we draw about the source?  

\subsection{Source of the polarized emission}

We expect that at wavelengths of $\lesssim 1.3$\,mm most of the flux originates
from a small region, probably less than a milliarcsecond ($\lesssim0.4$~pc)
across, centered close to the nucleus.  For an AGN with radio jets the mm
emission ``core'' is thought to be located somewhere in the approaching jet,
displaced from the black hole.  

In blazars, where the jet is closely aligned with our line of sight, the core
may be offset by thousands of Schwarzschild radii ($R_S$) from the black hole,
near the end of the zone where the jet is electromagnetically accelerated,
because this is where Doppler boosting is greatest.  For jets that are viewed
at a substantial angle, however, this model predicts that the core should be
close to the base of the jet \citep{Marscher2006}.  In M87, for example, where
the jet is inclined by $\sim$20$\arcdeg$ with respect to the line of sight,
VLBA observations by \citet{Hada2011} show that the 7\,mm radio core is
offset by only 14--23 $R_S$ from the black hole, while 1.3\,mm VLBI
observations appear to resolve the base of the jet, just 2.5--4~$R_S$ from the
black hole \citep{Doeleman2012}.  The jets in 3C\,84 are mildly relativistic
\mbox{(0.3$c$ -- 0.5$c$)} and are directed at an angle of roughly $30\arcdeg$
to $55\arcdeg$ to the line of sight \citep{Walker1994,Asada2006}, so here too
the offset of the core from the black hole may be small.

Variations in the polarization position angle presumably are caused by changes
in the magnetic field structure of the emitting region, possibly as the result
of shocks propagating along the jet similar to what is seen in blazars
\citep[e.g.,][]{Aller1999}.  Optical emission originates in these
same shocks, although from volumes that are much smaller, leading to faster
fluctuations in the optical position angles \citep{Jorstad2010}.  The rotation
of 3C\,84's optical and mm polarization position angles with time is
reminiscent of the systematic variations seen in BL\,Lac in late 2005.  In
BL\,Lac, this rotation was correlated with an optical, X-ray, and radio
outburst, and was attributed to a shock propagating along a helical magnetic
field in the jet \citep{Marscher2008}. 

\subsection{Location of the Faraday screen}

Where, then, is the Faraday screen?  Is it close to the nucleus, or far away in
the intracluster gas?  The rotation measure is given by \citep{Gardner1966}
$$RM = 8.1 \times 10^5 \int n_e\,{\bf B} \cdot d{\bf \ell} \ \ \ {\rm radians\
m^{-2}},$$ where $n_e$ is the thermal electron density in cm$^{-3}$, ${\bf B}$
is the magnetic field in gauss, and $d{\bf \ell}$ is the path length along the
direction of propagation in pc.  Only the component of the magnetic field along
the line of sight contributes; if the field is tangled, with many reversals
along the line of sight, the RM will be reduced.  

It is implausible that the Faraday rotation originates in the intracluster gas.
Typical RMs toward cooling flow clusters are in the range $10^3$ to
$10^4$~\rmsq\ \citep{Carilli2002}, similar to the RM of 7000~\rmsq\ measured by
\citet{Taylor2006} 15 mas (5 pc) from 3C\,84's nucleus.  The RM could be higher
if we happen to view the nucleus along the axis of one of the partially ionized
filaments that thread the intracluster gas surrounding NGC1275
\citep{Conselice2001}.  These filaments, \mbox{$\lesssim70$~pc} in diameter and
several kpc long, are stabilized by 10$^{-4}$ gauss magnetic fields
\citep{Fabian2008}.  If our line of sight to the nucleus passed precisely along
the axis of such a filament it could account for the measured RM, but such
perfect alignment is improbable. 

Probably the Faraday screen is close to the nucleus, within a parsec of the
emission core.  We cannot be certain whether the material in this screen is
being blown out from the black hole or is accreting onto it.  We consider these
two possibilities below.

\subsection{Faraday rotation in the jet boundary layer?}
\label{section:sheath}

Faraday rotation might originate in the sheath or boundary layer of the radio
jet, in plasma that is flowing outward from the black hole.  \citet{Zavala2004}
suggested such a geometry to explain the Faraday rotation measured in a sample
of 40 radio galaxies and quasars observed with the VLBA at wavelengths of 2~cm
to 3.6~cm.  Rotation measures were typically $10^3$ to $10^4$~\rmsq\ for the
radio cores in these sources.  This is comparable to the RM of about
7000~\rmsq\ measured in the 3C\,84 jet 5~pc from the nucleus by
\citet{Taylor2006} at wavelengths of 1.3 to 3.6\,cm.  The much higher RM that
we measure at 1.3\,mm might be explained if the mm emission originates closer
to the base of the jet and thus propagates through a denser zone of the
boundary layer.

In fact, an increase of RM at shorter wavelengths appears to be common in radio
jets.  In an AGN polarization survey, \citet{Jorstad2007} found that the RM
measured at mm wavelengths was greater than the RM measured at cm wavelengths
in 8 of 8 sources; a fit to these data gave $|{\rm RM}(\lambda)| =
\lambda^{-a}$, with $a = 1.8 \pm 0.5$.  This dependence can be explained by a
simple model in which the $\tau \sim 1$ surface is located at distance $d
\propto \lambda$ along the jet, and where the magnetic field, path length, and
electron density in the boundary layer scale as $d^{-1}$, $d$, and $d^{-2}$
respectively, giving $|{\rm RM}(\lambda)| \propto \lambda^{-2}$
\citep{Jorstad2007}.  

For 3C\,84, scaling the 1.3\,cm RM of 7000~\rmsq\ by $\lambda^{-2}$ gives RM
$\sim 7 \times 10^5$~\rmsq\ at 1.3\,mm, in good agreement with the measured
value.  We caution that this agreement may be a fortuitous coincidence.  The
model assumes that the cm emission originates 10 times farther from the nucleus
than does the mm emission.  In fact, however, the mm emission likely originates
within a few $\times$ 0.1~mas of the nucleus, whereas the cm RM was measured at
the tip of the jet 15 mas away, so the actual distance ratio is closer to 100.

\begin{figure}
\centering
\includegraphics[width=1.0\columnwidth, clip, trim=1.5cm 6.5cm 1.5cm 9cm]{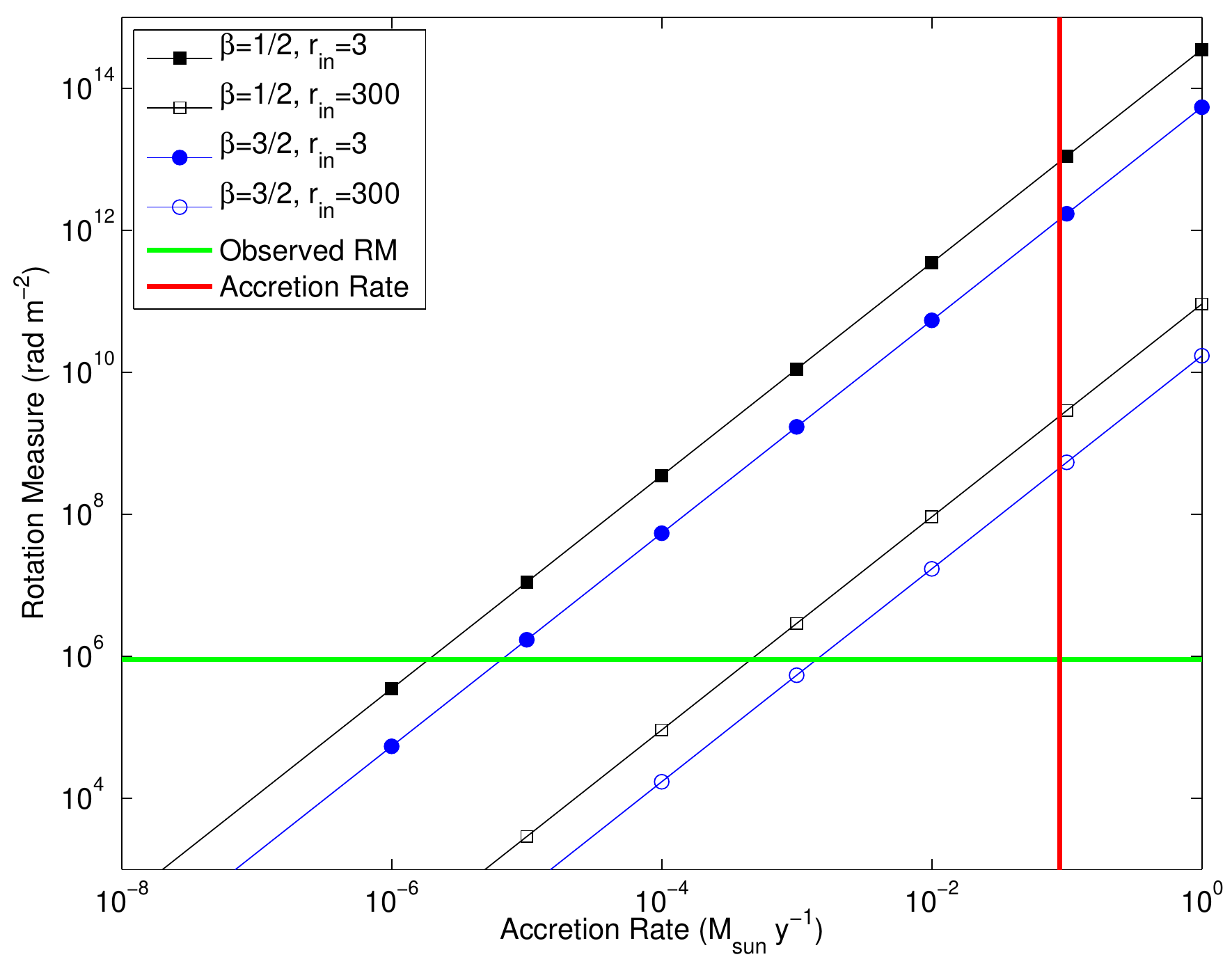}
\caption{Rotation measure vs. accretion rate for 3C\,84 predicted by
radiatively inefficient accretion flow models.  The RM depends on $\beta$, the
density power law index ($n(r) \propto r^{-\beta}$), and $r_{in}$, the radius
where the electrons become relativistic, given in units of the Schwarzschild
radius $R_S$.  The accretion rate is the mass inflow rate at $r_{in}$.  The
horizontal green line indicates the measured RM; the vertical red line
indicates the accretion rate estimated from the bolometric luminosity and a
10\% radiation efficency.  These spherically symmetric models fit the measured
RM only if an unrealistically large value for $r_{in}$ is assumed.}
\label{fig:rm}
\vspace{0.1in}
\end{figure}

\subsection{Faraday rotation in the accretion flow?}

We now consider the possibility that the Faraday rotation originates in the
accretion flow onto the black hole.  The RM, $\sim 9 \times 10^5$~\rmsq, is
among the largest ever detected.  However, it is striking for the fact that it
is not larger.  It is less than a factor of two greater than the RM observed
toward SgrA*, which is thought to originate in a radiatively inefficient
accretion flow (RIAF) surrounding the black hole \citep{Bower2003,Marrone2007}.
For SgrA* the RM constrains the accretion rate onto the black hole to be
$\lesssim10^{-7}~M_\odot\,y^{-1}$; Bondi accretion is excluded because it
requires an even higher RM.  Accretion onto the black hole in 3C\,84, on the
other hand, powers a massive outflow into the Perseus Cluster.  In 3C\,84 the
black hole mass is $8 \times 10^8\, M_\odot$ \citep{Scharwachter2013}, 2.5
orders of magnitude larger than SgrA*, and the total luminosity is $4 \times
10^{44}\ {\rm erg\, s^{-1}}$ \citep{Levinson1995}, 9 orders of magnitude
larger.  If the RM scales with the black hole mass or mass accretion rate, we
might expect it to be orders of magnitude larger in 3C\,84.

Accretion flow models fall into two classes.  RIAF models should be applicable
to sources with luminosities less than about 1\% of the Eddington luminosity
\citep{Narayan2012}; thin disk models \citep{Shakura1973} are more appropriate
for higher luminosity sources.  3C\,84's luminosity is about 0.4\% of its
Eddington luminosity of $\sim 10^{47}$~ergs~s$^{-1}$, so it is reasonable to
use RIAF models to predict its RM.


Following the formulation for the RM as a function of accretion rate for
spherical power-law accretion profiles in \citet{Marrone2006}\footnote{Note
that there is a typographical error in equation (9) of \citet{Marrone2006} --
the power law index for $r_{in}$ should be $-7/4$ \citep{Macquart2006}.}, we
calculated the RM for RIAF models with inner radii $r_{in}$ of 3 and 300
Schwarzschild radii $R_S$ (Figure~\ref{fig:rm}).  $R_S=8 \times 10^{-5}$ pc for
an $8 \times 10^8\, M_\odot$ black hole; $r_{in}$ is the radius at which the
accretion flow becomes relativistic, or is truncated for some other reason.  We
assume a very large outer radius, $r_{out} \sim 10^5\, R_S$, about 8~pc.  The
results are not sensitive to $r_{out}$ or to the density power-law index,
$\beta$, which selects between the limiting cases of advection-dominated
accretion flow (ADAF; $\beta = 3/2$) and convection-dominated accretion flow
(CDAF; $\beta = 1/2$) models.  

For $r_{in}=3 R_S$ the measured RM implies an accretion rate $\lesssim 10^{-6}
M_\odot\, {\rm y^{-1}}$.  This value is strongly inconsistent with the
accretion rate $\dot{M} \sim L/(0.1\,c^2) \sim 10^{-1} M_\odot\, {\rm y^{-1}}$
estimated from the bolometric luminosity and a radiative efficiency of 10\%; an
even higher accretion rate is required if the radiative efficiency is lower.
We can account for the RM in an ADAF context only if $r_{in} = 3000R_S \sim
0.2$~pc.  Even larger values are required for CDAF models.  But these inner
radii are much larger than any theoretical expectations.  If the polarized
radiation at 1.3\,mm originates close to the black hole, then either the
magnetic field in the accretion flow is much weaker than the equipartition
value assumed in the calculation, or the field is highly tangled, or the
accretion flow is disk-like rather than spherical.

Other observations suggest that material close to the nucleus of NGC1275 lies
in a disk that is tilted with respect to the line of sight. For example,
\citet{Scharwachter2013} model NIR observations of ionized species in the inner
$1.5''$ (50 pc) region as originating from a disk at an inclination angle of
45$\arcdeg$, with electron density $n_e\sim4\times10^3\,{\rm cm^{-3}}$ and
temperature $T_e\sim15000$~K. 

This disk is detectable on even smaller scales via radio free-free absorption
in multifrequency VLBA images \citep{Walker1994,Walker2000}.  Absorption is
seen against the N counterjet, but not toward the nucleus or the S jet.
Modeling suggests that it originates in a torus with $n_e \sim 10^4~{\rm
cm^{-3}}$, $T_e \sim 10^4$~K, and $L\sim 3$~pc \citep{Levinson1995}.  The
equipartition magnetic field in this gas is $B_{eq} = 4(\pi n_e kT)^{1/2} \sim
0.8$~mG.  If the polarized mm emission passed through 3~pc of this material,
the RM could be as large as $2 \times 10^7$~\rmsq, 20 times the measured value.
Probably line of sight to the mm core intercepts only a small fraction of this
material.  Since we do not know the exact location of the mm emission region
relative to the black hole, it is difficult to constrain the scale height of
the disk.

The absence of measurable free-free absorption toward the nucleus at 1.3\,cm
can be explained if the cm wavelength emission originates farther downstream in
the jet due to optical depth effects, as in the model described in
Section~\ref{section:sheath} above.  

The jet efficiency, $\eta_{\rm jet}$, defined as the ratio of the
jet power $P_{\rm jet}$ to the accretion power $\dot{M}_{BH}c^2$
onto the black hole, has been used to explore the mechanisms through which jets
are launched, as well as the role of black hole spin and magnetic fields
\citep[e.g.,][]{Nemmen2014}.  $P_{\rm jet}$ may be inferred from the
energetics of X-ray cavities excavated by the jets.  If one assumes that
accretion onto the black hole occurs at the Bondi rate, then it is typical to
infer jet efficiencies of a few percent \citep{Allen2006}.  Bondi accretion is
spherically symmetric inflow from the accretion radius $r_B = 2 G
M_{BH}/c^2_s$, where the sound speed $c_s$ is estimated from X-ray
observations of the gas temperature near the center of the galaxy.  For 3C\,84,
as in most radio galaxies, this temperature is in the range 0.5--3~keV
\citep{Fabian2006}, so the accretion radius is tens of parsecs.  Our RM results
and the free-free absorption data suggest, however, that accretion is disk-like
on scales smaller than $r_B$, which implies that estimates of
$\dot{M}_{BH}$ based on spherical Bondi or RIAF models may not be
valid for all sources, especially those near the transition between RIAF and
thin-disk accretion.

\subsection{Time variability}

On timescales of decades the emission from 3C\,84 varies dramatically, both in
the radio and in the $\gamma$-ray band; currently the source is brightening
rapidly \citep{Dutson2014}, suggesting increased fueling of the black hole.
Figure~1 in \citet{Dutson2014} shows that the 1.3\,mm flux density increased by
a factor of about 1.6 from mid-2011 to mid-2013. Over this same time span our
polarization measurements show no apparent systematic increase in the RM.  This
suggests that processes inside r$_{in}$ control accretion onto the black hole,
or that our line of sight to the mm core does not pass through the inner
accretion flow.  More precise measurements of the RM would be valuable to
search for variability caused by turbulence or patchiness in the accretion
flow, as in the SgrA* models of \citet{Pang2011}.  Or, if the polarized mm
emission originates in a hot spot moving outward along the radio jet, then
changes in the RM could be used to probe the structure of the accretion flow as
a function of radius.

\section{Summary}
\label{sec:summary}

Polarization observations with CARMA and the SMA show that radio source 3C\,84
is linearly polarized at wavelengths of 1.3\,mm and 0.9 mm.  The variation in
position angle with wavelength is consistent with Faraday rotation, with a
rotation measure of ($8.7 \pm 2.3$)$ \times 10^5$~\rmsq, among the largest ever
measured.  The fractional polarization was 1--2\% over most of the 2 years
spanned by these observations.  The rotation measure was stable within
$\pm$50\% over this period, even as the polarization position angle drifted
steadily toward more negative values, wrapping through a span of roughly 300
degrees. 

We argue that at mm wavelengths the linearly polarized radiation from 3C\,84
originates from the nucleus of the system, possibly within tens of
Schwarzschild radii of the black hole, and that the Faraday screen lies just in
front of the emission region.  It is uncertain whether the Faraday rotation
originates in in the boundary layer of the radio jet, or in the accretion flow
onto the black hole.  We investigated whether quasi-spherical radiatively
inefficient accretion flow (RIAF) models could explain the measured RM but
found that they overpredicted it by several orders of magnitude.  This suggests
that on scales of less than a parsec the accretion flow onto the black hole is
primarily disk-like rather than spheroidal.  The geometry of the disk
previously inferred from free-free absorption appears to be correct, with the
disk obscuring the counterjet and the innermost parts of the core.  

More highly inclined systems such as Centaurus A may exhibit even larger RMs.
Such sources would appear unpolarized in broadband observations.
Spectro-polarimetry at mm wavelengths with CARMA, SMA, and ALMA provides a
powerful tool to uncover the accretion flows in these systems.

\acknowledgments

Support for CARMA construction was derived from the states of California,
Illinois, and Maryland, the James S. McDonnell Foundation, the Gordon and Betty
Moore Foundation, the Kenneth T. and Eileen L. Norris Foundation, the
University of Chicago, the Associates of the California Institute of
Technology, and the National Science Foundation. Ongoing CARMA development and
operations are supported by the National Science Foundation under a cooperative
agreement, and by the CARMA partner universities.  DPM is supported by the
National Science Foundation through award AST-1207752.

The Submillimeter Array is a joint project between the Smithsonian
Astrophysical Observatory and the Academia Sinica Institute of Astronomy and
Astrophysics, and is funded by the Smithsonian Institution and the Academia
Sinica.

This study makes use of 43 GHz VLBA data from the VLBA-BU-BLAZAR Monitoring
Program (VBUBMP; http://www.bu.edu/blazars/VLBAproject.html), funded by NASA
through the Fermi Guest Investigator Program. The VLBA is an instrument of the
National Radio Astronomy Observatory. The National Radio Astronomy Observatory
is a facility of the National Science Foundation operated by Associated
Universities, Inc.

{\it Facilities:} \facility{CARMA}, \facility{SMA}, \facility{Perkins}.

\clearpage

\end{document}